# A quantum approach to the uniqueness of Reality(*)

by **Roland Omnès**

Laboratoire de Physique Théorique, Université de Paris XI  (Orsay)

*Abstract*

A brief review is given of the present state of an approach to consistency between basic quantum mechanics and a unique macroscopic reality, with no assumption of branching in the state of the universe. The main new idea consists in the recognition of local properties in the growth and transport of entanglement between a macroscopic measuring quantum system and a microscopic measured one. Moving waves of entanglement from the environment arise then and carry external phases, affecting significantly the state of the measuring device. These "predecoherence" waves perturb randomly the growth of other waves, which carry entanglement with the measured system. The outcome of these wave interactions could generate random fluctuations in the quantum probabilities of different measurement channels, which could lead in turn to a collapse mechanism satisfying Born's probability rule, according to earlier works by Nelson and Pearle.

A necessary randomness in the environment remains however unexplained and some suggestions regarding algorithmic complexity of the wave functions in a large quantum system e are made along that direction.

___________________________________________________-

## 1. Some aspects of the measurement problem

The measurement problem in quantum mechanics is well known and thoroughly discussed in many books (see particularly [1-4]). Its central difficulty is the existence of a unique experimental reality (at least at a macroscopic scale), because this uniqueness is (at least apparently) at variance with the quantum superposition principle.

Many attempts were made to obtain an answer. The Copenhagen school, after Bohr, brought out the idea of a (somewhat poorly defined) wave function collapse. Von Neumann, then later London and Bauer and also Wigner envisioned the consciousness of an observer as a source of consistency in a positivistic approach to human knowledge. Bell renewed the idea of hidden variables in a definite way, with the great advance of allowing experimental tests. Ghirardi, Rimini and Weber proposed an effect of spontaneous localization as an addition to quantum mechanics. There was also Bohm's proposal of a unique reality at the level of particles, this reality being parallel to the world of wave functions and guided by these functions. Several more proposals were also worth attention and the total number of these attempts witnesses by itself both the difficulty and the significance of the problem.

I shall however restrict this talk to the assumption according to which quantum mechanics could be essentially able to provide its own interpretation and reveal by itself from its own construction the origin of a unique reality[1].

There is only presently one theory of this type, which is Everett's formulation of multiple realities [5]. When completed with Von Neumann's chains of measuring devices [6], this interpretation brings out strict correlations of different observers' consciousness regarding the data of which they are aware. This approach is therefore consistent from the standpoint of positivism [7].

One may be reluctant however about the philosophical consequences and the mathematical background of this interpretation. Philosophically, one may be worried by the extreme holism of a theory where a huge number of distinct universes emerge from each one of many tiny events, which happens to act like would do a quantum measurement. This is something new in philosophy but there is also something puzzling when one must pay such drastic metaphysical requirements as the price for getting a positivistic satisfaction.

There are also mathematical difficulties in the assumption of a wave function for the universe. It means first of all a gigantic extension of the validity domain for the quantum axioms, with the sole benefit of naming at last the universe as a system that can be presumed isolated. This extension requires moreover a tremendous precision in the algorithmic content of this wave function, if it can account for all the states and properties of every atom or particle in the universe as well as the detailed state of every subsystem, however large or small. This means that one is not only assuming the existence of many branches of the universe but also a level of exactness in quantum mechanics exceeding any conceivable empirical justification. The necessary precision of quantum laws would have then to involve many powers of powers of ten[2], whereas it is presently of order $10^{-12}$ according to the best experiments.

Since this meeting is devoted to Turing's legacy, one is also led unavoidably to wonder what consequences could follow from the high algorithmic content of a wave function of the universe. Can one take for granted that it would not induce some kind of algorithmic randomness, in which case any reliance on a deterministic unitary evolution of the universe would be jeopardized?.

---

[1] This is of course Einstein's question regarding the completeness or incompleteness of
[2] In quantum field theory, these requirements become still more stringent, and much more so again when gravitation is considered, for instance through string theory.

From there on, anyway, one will directly proceed to a search for some conditions under which a macroscopic uniqueness could be generated in a measurement, when the measuring device is macroscopic. Although this perspective is often considered as impossible without a violation of the quantum principles, the perspective of a new opening will be proposed and partly developed.

Leaving from there on aside more introduction dealing with the usual arguing against arguments, so familiar in measurement theory, we shall now go directly to the matter at hand, at least as it stands presently.

**2. About entanglement**

It is well known that one of the main impediments against compatibility of the quantum rules with a unique reality is entanglement [8, 9]. In spite of valuable research [10] (inspired by the perspective or dream of quantum computers), too little attention has been paid in my opinion to local properties of entanglement, which certainly exist and will be now considered.

Abstractly, entanglement is concerned with an isolated system (say $S$), which is made of two subsystems (say $A$ and $B$). Initially, $A$ and $B$ are independent (*i.e.*, isolated from each other). Mathematically, this means that the state vector (or wave function) $|S>$ of the whole system is a (tensor) product $|A>\otimes|B>$ of two states of the subsystems[3]. For some reason, an interaction occurs between the two subsystems at some time[4], then stops or is complete after some more time. These are at least the circumstances under which von Neumann encountered entanglement [6] and how Schrödinger defined it [11]. The essential point is that, when $A$ and $B$ cease to interact, they have no more mutually independent wave functions $|A>$ and $|B>$ and the wave function $|S>$ is no more a product. In other words, even an isolated system bears the mark of another system with which it interacted in the past and it never recovers the virginity of having its own wave function.

This is both abstract and startling and Schrödinger made it plain with his famous example of a cat [12]. Leaving some details aside, the system $A$ is then a radioactive source and $B$ is a cat. Initially, they are independent. The effect putting them eventually into interaction is internal to $A$ and consists in its possible radioactive decay. The entangled outcome is the famous expression of the state of the source-cat system as a superposition of products:

$$|A, \text{source intact}>\otimes|B, \text{cat alive}> + |A, \text{decayed source}>\otimes|B, \text{dead cat}> \quad (1)$$

It will be convenient to use an example where the system $B$ is macroscopic and consists of a Geiger counter containing a gas of argon atoms and the (measured) system $A$ is an energetic charged particle crossing this counter along a straight-line trajectory. There is no superposition as in Equation (1) in that case but the final state of the counter has been anyway modified after a complete interaction with $A$ (whether the particle finally came out of the counter or was stopped). Just for curiosity, one will look at what happened when $A$ was crossing $B$, our purpose being to look later at the generation of an entangled state and not only at its final expression. Our main interest will not only be to follow this generation as time goes on, but how it proceeds progressively in space and time. In other words, one is interested in the locality, growth and transport of entanglement during its generation.

---

[3] More simply, one can say that the wave function of $S$ is the product of two wave functions for $A$ and $B$, depending on different variables (for instance the positions of atoms in two different objects).

[4] For instance, the two systems were spatially separated initially and their motion brought them into contact, or one of them is microscopic and penetrates into the other one, which is macroscopic.

This is a somewhat technical question but it can be explained rather simply if its discussion is split into four steps, as follow.

*Step 1: Feynman histories*

Quantum mechanics can be expressed by means of Feynman histories and, in the present case, a history involves completely unconstrained motions for all the argon atoms and for the charged particle between time 0 when the particle enters in the counter and some later time $t$ A classical action function $S$ is associated with this history. When considering a specific argon atom $a$, one may split the set of histories into two subsets, according whether or not $a$ became "connected" with the charged particle $A$ before time $t$. To say that $a$ is not connected with $A$ means that $a$ did neither interact directly with $A$ nor with any atom belonging to a chain of atoms having had mutual interactions and including at least one atom that interacted with $A$. This is shown in Figure 1 where the weird variation of an atom position during its motion is symbolized by a straight horizontal line and the interaction by double vertical lines. The atom, denoted by $a$ in this figure, is not connected with the charged particle $A$, whose history is shown by a heavy line[5]

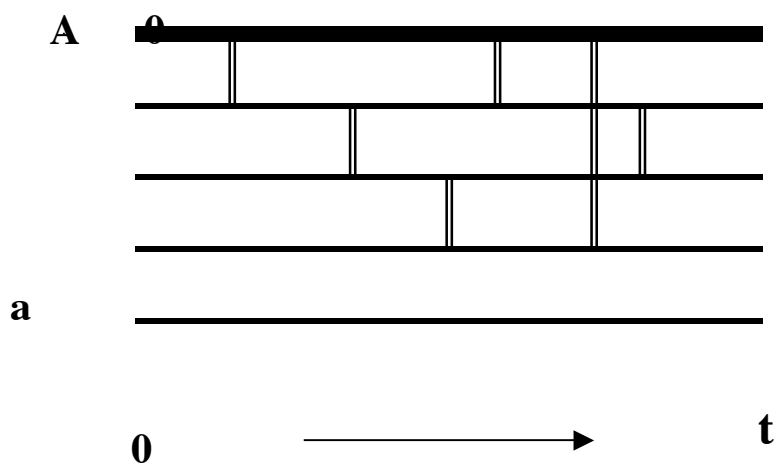

Figure 1: The topology of connectedness in a Feynman graph

*Step 2: Wave functions*

The connection in Feynman histories can be extended to wider questions, including a description of entanglement. As a matter of fact, Feynman histories were used here only for introducing clearly, by a drawing, this topological property of graphs. But there are other useful graphs in quantum mechanics, particularly the Feynman graphs expressing a perturbation expansion for the time evolution of a wave function, and the same idea can be applied to them[6].

---

[5] This symbolic representation is meaningful, since connection is only a topological property.

It becomes clear, as one proceeds, that the corresponding connectedness is closely linked with kinetic theory, whose relation with quantum mechanics is notoriously difficult. Here, I propose to relate these two aspects from the start through different topologies for wave functions. There is on one hand the algebraic topology of the quantum Hilbert space and, on the other hand, the topology of connections in Feynman graphs. The two of them will be conveniently joined within a common algebraic framework.

A Feynman graph for a contribution of a definite term in perturbation theory to the evolution operator of a system of particles has the same topological properties as a Feynman history and both show the same possibilities of connection: If one considers for instance two argon atoms, denoted by indices $n$ and $n'$ and interacting at some time, the definition of connectedness implies that if both atoms were initially connected with the charged particle $A$, they remain so after an interaction. The same conservation of the properties of connection remains true when both atoms were not connected with $A$. However, when one atom is connected and the other one is not, the two of them are connected after interaction.

There is a simple algebraic way for expressing these properties of connectedness with $A$ by means of the interaction potential $V_{nn'}$ between the atoms $n$ and $n'$: One introduces a "connectedness index" for each atom: this index is 1 if atom $n$ is connected and 0 if it is not connected. Proceeding similarly for atom $n'$; one gets four possible combinations of different connections with $A$ before or after a $n$-$n'$ interaction. One can then write down easily a 4×4 matrix $O_{nn'}$ expressing the changes in connectedness under an interaction and insure a bookkeeping of connectedness by a replacement of $V_{nn}$ in the Schrödinger equation by the matrix $V_{nn} O_{nn'}$ [13].

Under this change, the wave function of the system $A + B$, which showed no connection before the entry of the charged particle $A$ into the Geiger counter $B$, becomes at a later time $t$ a sum of component wave functions in which every term exhibits a definite connection for every atom. The total wave function still evolves nevertheless under the standard Schrödinger equation and the evolution of connectedness is only seen in the components, which obey altogether a refined well-defined evolution equation[7].

*Step 3*: *Quantum fields*

I shall be still more allusive when coming to the next level, because each step becomes more technical. Essentially, one relies then on the so-called "second quantization" of atomic physics, which deals with quantum fields. One introduces a field $\varphi(x)$ to describe the atoms (and the charged particle $A$), where the notation $x$ involves the position of an atom and also spin indices. A direct consequence of Step 2 is the existence of two fields $\varphi_1(x)$ and $\varphi_0(x)$, the first one being associated with connected states of atoms and the second with disconnected states.

---

[6] This kind on connectedness is not new. It is used in the foundations of statistical mechanics and in Faddeev's theory for multiple scattering. As a "clustering property", it was used by Steven Weinberg as one of the main foundations of quantum field theory.

[7] Algebraically, this means that the Hilbert space of $B$ is meant as the direct sum of linear spaces expressing connectedness with $A$. These spaces are not orthogonal however, which means that connectedness cannot be expressed by means of an observable. It is not a "physical property" as was meant by Von Neumann [5] and this conceptual change could explain much of the difficulty in understanding measurements.

This straightforward description has several consequences:. It deals more easily with indistinguishable atoms than in Step 2. It expresses also more directly the average number of atoms that are connected in some space region through a probability density $\rho_1(x) = \varphi_1^\dagger(x)\varphi_1(x)$ for connected atoms with $A$ and a similar density for disconnected atoms. From these densities, one can get a probability $p_1$ for the atoms to be connected in some space region (and $p_0$ for them being disconnected). The sum $p_1 + p_0$ is not however strictly equal to 1, but very close to 1 for a macroscopic region, essentially because of the large number of atoms in the detector $B$. Moreover, the approach through quantum field theory provides a wider generality allowing to account for instance for ionization and creation of free electrons when the particle $A$ interacts with neutral argon atoms, as well as the emission of photons from the decay of an atomic excited state.

As a matter of fact, every theoretical tool entering in the quantum description of a detector is expressible by means of this quantum field version of connectedness[8]. It shows plainly the local repartition of entanglement before its completion as well its growth and transport, which we shall now consider at a macroscopic scale.

*Step 4*: *Macroscopic entanglement transport*

We thus arrived at a representation where entanglement appears as a topological property of the evolution of atomic states, which is carried by the motion of atoms and transmitted by them through interactions, or more precisely through collisions in the present example. This conception of entanglement is strongly reminiscent of a transport process (such as heat conduction and the conductivity of heat or of electric charge) and the next step will be therefore to look at these transport properties..

A significant difference with more familiar transport processes is however a *contagious* character of entanglement (an atom can catch connectedness from an already connected atom, like people catch flue when they come close). This is different from the conservation of energy or electric charge, where there is a share rather than contagion. There must be therefore a local growth of entanglement where it has been created or already exists, together with transport from a region where there is much entanglement towards a poorer region. Ultimately, after a long enough time, entanglement should become total everywhere and hold for the whole state of the Geiger counter.

Some measure of entanglement is needed to give more substance to these ideas. This measure must be local, at least when considered as a density of entanglement from a macroscopic standpoint. To this end, one denotes by $f_1(x, t)$ the average probability for connectedness of atomic states in a small macroscopic region around a space point $x$ at time $t$. Similarly, $f_0(x, t)$ will denote the corresponding measure for disconnected states, and one assumes again that the sum $f_0(x, t) + f_1(x, t)$ is very close to 1.

The transition from quantum fundamental effects to the kinetic behavior at a large scale is however among the trickiest points in theoretical physics, although kinetic theory is usually valid at a macroscopic scale. This remark, although only methodological, will be our sole justification for making a final step, notwithstanding that entanglement is an extreme paradigm of pure quantum mechanics.

---

[8] To summarize the algebra of this description for a system of $N$ identical atoms, a pure Schrödinger wave function belongs to a Hilbert space $E$, made of symmetric square summable functions of $N$ variables. A time-dependent description of entanglement consists in writing such a function $\psi$ as a sum $\Sigma_q \psi_q$ of wave functions expressing a repartition of the atoms between $q$ atoms connected with $A$ and $N-q$ unconnected. One may consider a component $\psi_q$ of $\psi$ as belonging to a Hilbert space $E_q$, isomorphic to $E$, but there is no property of orthogonality holding between different $E_q$'s. of which the total number is $2^N$.

One assumes the gas in the Geiger counter at thermal equilibrium. If connectedness were conserved under a collision, entanglement would be a conserved quantity and the random walk of a connected atom would result from collisions with other atoms, whether they are themselves connected or not. The quantity $f_1(x, t)$ would then obey a diffusion equation. Because however of the contagion of entanglement, there is also an increase in $f_1$ from collisions with contagion and one can show that the resulting evolution equation is

$$\partial f_1 / \partial t = f_1 f_0 + (1/6)\nabla^2 f_1, \qquad (.3)$$

where the mean free path and mean free time of atoms were taken as units.

Because $f_0 = 1 - f_1$, this equation is nonlinear. The writing of boundary conditions is tricky but one can get a hint about them from a remarkable property of entanglement: When two atoms collide and contribute to the contagion of entanglement, both of them are entangled in the final state. If we define for convenience a space axis along some direction of space $x$, a remarkable property comes out from the indistinguishable character of the two atoms, namely that, *on average*, one atom goes out with a velocity component $v$ along the $+x$ direction and the other atom with an opposite velocity component. One can show moreover that this average velocity component $v$ is equal to the velocity of sound. This property of contagion implies that in the present case, the propagation of entanglement proceeds at the velocity of sound. This remark can be used to guess an answer for the problem of boundary conditions for equation (2.2), namely that the domain where $f_1$ is non-vanishing at a time $t$ is bounded by (formal) sound waves, which would have been emitted by particle $A$ along its track when it moved in the Geiger counter.

This mixture of physics with mathematics would need more rigorous improvements but it leads to rather sensible predictions: The growth of entanglement should occur behind a rather sharp front moving at the velocity of sound. This behavior is sensible since nonlinearity is known necessary for the spontaneous generation of a wave front [14]. As for the shape of this front as it results from computation, it is shown in Figure 2..

These properties are presumably general and no case of actual measurement seems inaccessible to this type of approach, at least in principle. Some adaptation would be of course necessary in different circumstances, to account for instance for actual ionization cascades in a Geiger counter or a wire chamber. As for macroscopic signals, which can act as the "pointers" in textbooks (such as a cascade of ions for instance), they cannot occur sooner than local entanglement, whose growth and transport appear therefore as the most significant characters in a measurement.

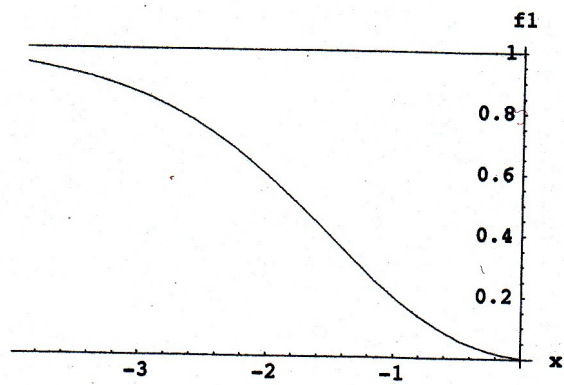

Figure 2: An entanglement wave, showing the local probability of entanglement behind a wave front (at a time when the front is located at x = 0). The unit of distance is an atomic mean free path.

*Superposition of states*

The discussion has been restricted till now to a case where the measurement is in some sense completely predictable, since the initial state of the incoming particle *A* was supposed to cross with certainty the Geiger counter *B*. The case when *A* misses *B* would be trivial and there would be no more entanglement at the end than before, *i.e.*, none.

The real problems of measurement theory occur when there is a superposition in the initial state of *A*, for instance

$$c_1 \,|\,A \text{ must enter into } B > + c_2 \,|\,A \text{ must miss } B > \qquad (4)$$

Nothing significant or new is found in that case when compared to the description of measurement theory in textbooks, *at least as long as one assumes the counter isolated*. To get more, one must turn to the effects of an environment.

## 3. Predecoherence

Decoherence is a well-known effect forbidding, or rather damping very rapidly quantum interferences at a macroscopic level[9]. It is extremely efficient since for instance the exposition of a pointer with an area of 1 centimeter square to an ordinary atmosphere divides by a factor of 2 the intensity of possible interferences after a few $10^{-24}$ seconds! This amazing property does not change however drastically the measurement problem since the various measurement channels have still conserved their initial probabilities after the measurement (such a probability is for instance the square modulus of a complex quantities $c_1$ or $c_2$ in the sum (3))

---

[9] One often says that because of decoherence, Schrödinger's cat is no more dead and alive but becomes "either dead or alive" (whatever this means in the mind of somebody hearing or reading this sentence).

The origin of decoherence is closely similar to the origin of entanglement and this point was particularly emphasized by Zeh and collaborators [3]. In the case of the pointer we just mentioned, the microscopic cause of decoherence lies in the collisions of external atmospheric molecules on the pointer. Every such collision is much similar to the effects we already described concerning the effect of particle *A* in the detector *B*: it brings out a connectedness (an entanglement) between the states of the pointer and outgoing molecules of air. If the system under consideration consists of the detector and the environment and if it can be considered as isolated (eventually by extending the environment as far as the whole universe...), nothing essential is changed when one compares this with the case when *B* was isolated: The various measurement channels are still conserved as well as their associated probabilities. Decoherence appears as a practical consequence of our empirical ignorance of the detailed quantum state of the atmosphere together with a necessary restriction of an observer's information to a "reduced state" of the pointer [10]

Not much more can be added regarding decoherence and its positivistic status, but something else is worth attention, which we may call *predecoherence*. Whereas decoherence is a physical process, predecoherence refers to the physical situation at some time *t* of the measuring system under the same interactions with the same environment. It is not necessarily restricted to a measuring device but also present in any non-isolated macroscopic system, such as a clock for instance.

The essential point is that every interaction with an external air molecule *M* generates a wave of entanglement with *M*, which moves in *B* at the velocity of sound, until it has crossed the whole of *B* and is afterwards acquired and ineffective.

As a matter of fact, this is mostly trivial and only means that a tremendous number of entanglement events with external molecules occur continuously in the counter *B*. Each interaction with an external air molecule *M* generates a wave of entanglement with *M*, which moves inside *B* at the velocity of sound until it crossed the whole of *B*. Such a wave has no special physical effect since it carries only a topological property of entanglement. Behind a wave front, all the argon atoms in the counter are entangled with *M* and they are not entangled for an atom before the front,. As shown on Figure 2, the transition takes place on the front within a region having a width of a few times an atomic mean free path[11] .

So, what is the matter? Not much for physics, at this point of the discussion, but there is at least one point of concerning an argument that is often made to restrict the extension of an environment: This is impossible and one can never fix an environment as finite and neglect what happens far away on its boundary, because there is always an environment of the environment, which always modifies its wave functions at every place through moving waves that were born long ago and far away. Since there is never complete isolation for the quantum behavior of a macroscopic system in the universe, it would seem therefore that one is again pushed back to some sort of Everett framework.

**4. The case of a random environment and the problem of collapse**

Before going further and since the method in science is a part of philosophy and has a place in an assembly of philosophers of science, I wish to explain the leading trend of the research I am proposing now[12]. As already said, the basic principles of quantum mechanics are considered as granted, at least as long as the properties under study, for the system under study, can be shown insensitive to the assumption of isolation. On the other hand, our almost complete ignorance of the wave functions of a macroscopic system and of their unknown algorithmic properties should be acknowledged. This last statement becomes obviously a great deal stronger when the "system" is the universe. For these reasons, I entertain personally a strong diffidence against some arguments intending to "prove" that a compatibility of a unique reality with the quantum principles is impossible [9][13].

---

[10] The reduced density matrix providing this information (and nothing more) is obtained through a trace operation over everything except the state of the measured particle *A* and the position of the pointer.
[11] about $10^{-5}$ cm under standard conditions of pressure and temperature.

An opposite miscalculation would be to hope for getting in one shot the answer to a great problem. This is why I do not attempt here to propose a full-fledged theory but only a search for hints. As for these hints, they should better rely on something that was left previously unnoticed or underestimated and I dare presume that the growth and transport properties of entanglement could be such a hint.

From there on, I shall make a strong assumption, which is that *the collisions of external molecules on the Geiger counter B can be considered as incoherent and random events*[14]. We shall later return to this postulate and only consider presently what the consequences could be.

Beforehand, let one contemplate predecoherence. Each external collision by a molecule brings out its own entanglement wave carrying an arbitrary phase. The number of these waves is extremely large but easily estimated because one knows their frequency of production and how much time they spend for crossing *B*. They contribute to thermal disorder but have no special consequence otherwise in ordinary conditions (in the state of a clock, for instance).

The permanence and randomness of predecoherence in *B* become significant on the other hand when a quantum measurement of the charged particle *A* occurs. In order to see better what this incoherence means, we can take for a comparison the case of an isolated system. We saw that in that case the growth of entanglement between *A* and *B* proceeds through an increase of connection with *A* of the atoms in a wave function of *B*, in both channels 1 and 2[15]. Under this increase in connection, a Schrödinger wave function $\psi_1$ of *B* in Channel 1 (for instance) becomes a sum $\Sigma_q \psi_{1q}$ of components with different connections with *A*. All these functions have a in common a unique arbitrary phase, whatever the measurement channel and the amount of connectedness. As a consequence, coherence holds in every aspect of the formalism when the system *A+B* is isolated.

When the measuring device *B* is not isolated and subject to predecoherence, the situation is quite different. A tremendous number of arbitrary phases are then present in *B*, behind different fronts of entanglement with outgoing external molecules, these fronts moving at the velocity of sound. The problem of finding what happens actually in these conditions is apparently new in physics and preliminary investigations indicate a high complexity[16].

---

[12] As a matter of fact, this spirit did not precede the research but came as a lesson from many blunders by the author.

[13] The devil often hides its trumps in ambiguous or hidden assumptions in this kind of argument. One may recall as such a case how well the impossibility of anti-ferromagnetism had been proved before its experimental discovery.

[14] Incoherence is meant here in the quantum sense of an arbitrary relative phase for the incoming wave functions of two distinct molecules. This assumption could need refinement in the case of two molecules arriving at almost the same place and almost the same place

[15] We suppose these two channels associated with two distinct tracks of *A* in *B*. The case of Equation (3) where *A* misses *B* in Channel 2 requires an adaptation.

Anyhow, some aspects of the evolutions are clear: The main question of interest is of course the possibility of fluctuations in the probabilities $p_1$ and $p_2$ of the two channels. When an atomic state $a$, belonging to Channel 1, entangled with $A$, collides with a non-entangled atomic state $a'$, the collision process is unitary *if and only if* $a$ and $a'$ share entanglement with exactly the same outgoing molecules in the environment. This condition is so restrictive that, in practice, the proportion of unitary interactions is small and most collisions are non-unitary. One can expect therefore fluctuations in the transition from non-entanglement to entanglement with $A$ of atomic states such as $a'$, because then the collisions of atoms occur at random in addition to being non-unitary. Some fluctuations in $p_1$ and $p_2$ could then result from the random occurrence of collisions and from the high complexity of topologically connected components in the global situation of $B$[17].

These processes are so involved that I have no pretense of understanding them thoroughly. I do not even claim that there are actually fluctuations in $p_1$ and $p_2$ since there is no cogent proof, but it would be certainly a challenge to prove on the contrary that there are no fluctuations. One can at least establish a general form for correlation functions, which is :

$$<(\delta p_1)^2> = <(\delta p_2)^2> = - <\delta p_1 \delta p_2> = K p_1 p_2 \delta t \qquad (5)$$

for the fluctuations $\delta p_j$ during a short time interval $\delta t$. The coefficient $K$ is the inverse of a "collapse time scale". Estimates for the value of this time look quite encouraging, but their fundamental justification remains problematic because it encounters difficult and fascinating new problems (as well as older ones) in the relations between quantum mechanics and kinetics

Anyway, a strong incentive for pushing investigations further is an *a priori* knowledge of the consequences of a finite value for the coefficient: According to a theorem by Philip Pëarle [15], fluctuations obeying (5) would then lead necessarily to a final result where one channel probability, either $p_1$ or $p_2$ in our example, has become equal to one and the other one to zero 0. This outcome, which amounts essentially to the emergence of a unique macroscopic reality at the end of the measurement, is a random event whose randomness is controlled by predecoherence. In other words, the randomness of prior interactions of the detector with its environment is responsible for a random collapse. The beauty of the theorem is another conclusions, which is that the (predecoherent) probabilities for different outcomes of collapse coincide with Born's probability rule for quantum measurements.

The meaning of this significant theorem by Pearle, which is an elaborate version of the gambler's ruin theorem by Huygens, can be illustrated in a simple way, worth mentioning: Let one consider the two measurement channels 1 and 2 as two gamblers and say for instance that the squared modulus of $c_1$ in (4) is a measure of the hope of Channel 1 for reaching reality (this is in some sense the definition of Born's rule). After every short interval of time $\delta t$, the two channels exchange an amount of hope $\pm \delta p$ satisfying (5) and the hope of Channel 1 increases or decreases by this amount. Finally, inevitably, there comes a last step in which one last hope, either of Channel 1 or 2, vanishes forever. If Channel 1 turns out to be the winner, its existence is then indisputable and one can say that its macroscopic features become real. From there, it is a matter of mathematics to show that the probability for the

---

[16] This is no surprise since the difficulties arise from the well-known estrangement between quantum effects and kinetic effects;ρρ

[17] Global entanglement with Channel 1 is shown for instance in the matrix $< 1 \,|\rho_{A+B}|\, 1 >$ having trace $p_1$, but there is no unitary property insuring that this trace is conserved.

coming of Channel 1 to Reality is equal to its initial hope. One can also compute the average time for reaching this achievement (*i.e.*, the duration of collapse), which turns out to be of the order of $K^{-1}$ (hence the importance of finding the value of $K$)

Another point worth mentioning is the consistency of this process of collapse with the non-separability of quantum mechanics, in spite of locality of the mechanisms at work [18]

We now reach a conclusion, which holds in three points: The first one is that we are still far from a cogent explanation for an emergence of the uniqueness of reality from quantum theory. We have got only hints but, on the other hand, these hints look reasonably suggestive. Much more work is certainly needed and it will be certainly demanding, since the key remains hold in the notoriously difficult frontier bush between fundamental quantum mechanics and kinetic physics[19]. This location in a little explored field of knowledge could explain by the way why this possible key went unnoticed and why one could presume that a consistency of basic quantum theory with collapse were impossible. I believe that this mental barrier should be removed, the more so since much could be learned also from an exploration of this frontier of physics.

The second point brings us back to algorithmic complexity. The partial results in the present last section rested on a strong assumption, namely incoherence and randomness of the interactions with a nearby environment. We saw that there is no frontier to an environment so that we must inquire into the properties of wave functions for a very large system, if not the full universe. What is then the *degree* of complexity and even of definiteness of such functions? What do we know or what can we learn about them? Do they hold the key for the emergence and existence of reality?

The third and last point is a summary of the present investigation: 1. There is no limit to an environment and the universe is so wide that its global quantum evolution is algorithmically complex (in a sense that must still be made explicit). 2. A locally well-defined system (such as a clock or a wire chamber for instance) is permanently in a predecoherent state where a host of entanglement waves with the environment run randomly through it. 3. A quantum measurement would also yield a generation, growth and transport of entanglement waves with the different states of a microscopic measured system, but only for a finite time. These various channels remain present thereafter in the ideal case of isolation. 4. When there is no isolation, predecoherent random waves break down the coherent unitary evolution the various channels, which become unstable for the ultimate benefit of a unique one, whose existence becomes a feature of Reality.

Finally, although this general description of collapse should be considered only as a conjecture, I believe that some of the problems it raises are now sufficiently well-defined and the general approach new enough to warrant independent and deeper investigations.

---

[18] This result was obtained during exchanges with Bernard d'Espagnat.
[19] The local properties of entanglement in Section 2 belong to this frontier.